\documentclass[floatfix,10pt,aps,prb,twocolumn,showpacs,a4paper, superscriptaddress]{revtex4-2}
\usepackage{amsmath}
\usepackage{amssymb}
\usepackage{graphicx}

\usepackage{multirow}
\usepackage{array}

\usepackage{booktabs}

\usepackage{hyperref}

\begin{document}


\title{Optimizing Supercell Structures for Heisenberg Exchange Interaction Calculations} 


\author{Mojtaba Alaei}
\email{m.alaei@skoltech.ru, m.alaei@iut.ac.ir}
\affiliation{Skolkovo Institute of Science and Technology, 121205, Bolshoy Boulevard 30, bld. 1, Moscow, Russia}
\affiliation{Department of Physics, Isfahan University of Technology, Isfahan 84156-83111, Iran.}
\author{Artem R. Oganov}
\affiliation{Skolkovo Institute of Science and Technology, 121205, Bolshoy Boulevard 30, bld. 1, Moscow, Russia}


\date{\today}

\begin{abstract}
In this paper, we introduce an efficient, linear algebra-based method for optimizing supercell selection to determine Heisenberg exchange parameters from DFT calculations. 
A widely used approach for deriving these parameters involves mapping DFT energies from various magnetic configurations within a supercell to the Heisenberg Hamiltonian. 
However, periodic boundary conditions in crystals limit the number of exchange parameters that can be extracted. 
To identify supercells that allow for more exchange parameters, we generate all possible supercell sizes within a specified range and apply null space analysis 
to the coefficient matrix derived from mapping DFT results to the Heisenberg Hamiltonian. By selecting optimal supercells, we significantly reduce computational time 
and resource consumption. This method, which involves generating and analyzing supercells before performing DFT calculations, 
has demonstrated a reduction in computational costs by 1-2 orders of magnitude in many cases.
\end{abstract}


\maketitle

Predicting material properties using theoretical and computational methods is appealing because it reduces the cost of designing materials for specific functions. 
However, predicting magnetic properties poses unique challenges due to both theoretical and computational difficulties. 
Theoretically, approximations in electronic structure calculations, especially in density functional theory, often result in inaccurate estimates of electron exchange-correlation energy, 
leading to incorrect predictions in magnetic systems~\cite{Alaei2023}. 
Even advanced techniques such as continuum quantum Monte Carlo can be unreliable for comparing magnetic states due to the fixed-node approximation~\cite{Needs2010}. 
Computational challenges include long run times and difficulties in achieving electronic convergence, particularly when using large supercells. 
Therefore, finding an appropriate supercell with fewer atoms can reduce computational cost and may even enable the use of more refined ab initio methods~\cite{CCSD_nature}, 
such as coupled cluster singles and doubles (CCSD)~\cite{CCSD} through density matrix embedding theory (DMET)~\cite{DMET1,DMET2}.

One common approach for obtaining thermodynamic magnetic properties is to derive a spin model Hamiltonian from electronic structure results, 
which can then be used to calculate such properties as transition temperatures. There are two main methods for deriving spin models: 
one involves explicitly calculating interatomic exchange interactions using Green's function methods~\cite{Lichtenstein2023}, 
while the other maps the total energies of different magnetic configurations from electronic structure calculations onto the spin model. 
This paper focuses on the latter method, which is applicable to any electronic structure calculation that determines total energy through variational methods.

The main interaction in a spin model Hamiltonian is the Heisenberg exchange interaction, 
so we assume the Hamiltonian includes only the Heisenberg term. 
However, including other interactions, such as the Dzyaloshinskii-Moriya interaction, does not impact the method for determining the optimal supercell presented in this paper. 
In the mapping method~\cite{Alaei2015}, the total energies of various magnetic configurations are used to extract the exchange interactions $J_{i,j}$ between magnetic moments at sites 
$i$ and $j$ in the Heisenberg term, given by $-\frac{1}{2} \sum_{i,j} J_{i,j} \hat{\mathbf{S}}_i \cdot \hat{\mathbf{S}}_j$, 
where $\hat{\mathbf{S}}_i$ and $\hat{\mathbf{S}}_j$ represent the normalized magnetic moment vectors.

For collinear magnetic configurations, where the spin vectors $\hat{\mathbf{S}}_i$ and $\hat{\mathbf{S}}_j$ simplify to a single value of $\pm 1$, the mapping equation reduces to:
\begin{equation}
\label{eq:hes}
E_k = \sum_i^{m} \alpha_{k,i} J_i + c_0
\end{equation}
Here, $k$ denotes the $k$-th magnetic configuration, $J_i$ represents the exchange interaction for the $i$-th nearest neighbor, and $E_k$ 
is the total energy obtained from electronic structure calculations for that magnetic configuration. The term $c_0$ is a constant. 
The coefficients $\alpha_{k,i}$ and $c_0$ form a matrix $\mathbb{A}_{n \times m+1}$, where $n$ is the number of different magnetic configurations. 
From this matrix, one can determine, for a given supercell, up to which neighbor the exchange interactions can be reliably calculated~\cite{Alaei2023}.

\begin{figure}
\includegraphics[scale=0.450]{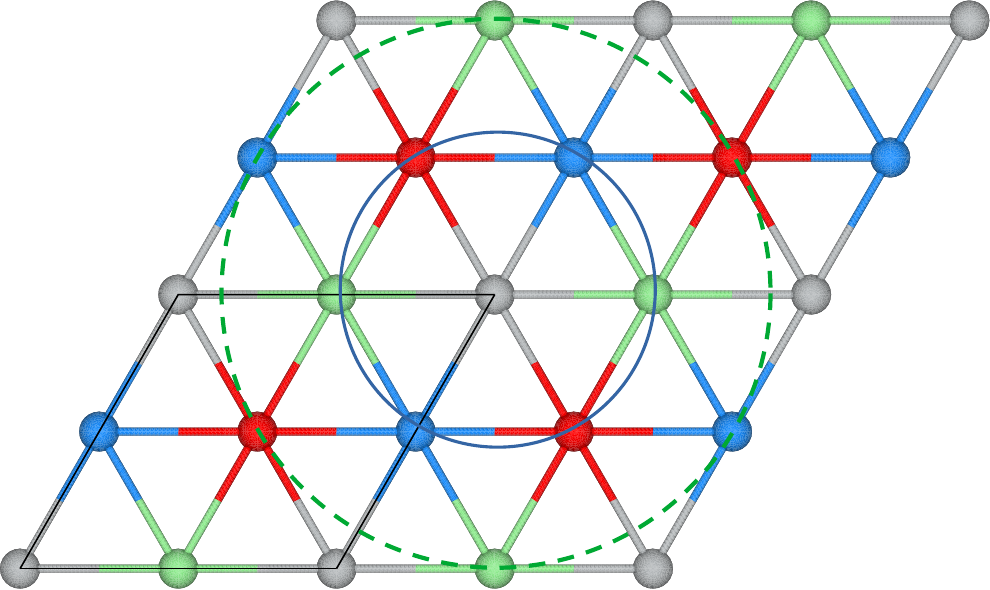}
\caption{
The figure illustrates the effect of periodic boundary conditions on Heisenberg exchange calculations. 
It shows a $2 \times 2$ supercell of a hexagonal monolayer, where each of the four atoms in the supercell is represented by a distinct color, 
indicating different magnetic directions. 
As shown, the colors for the first nearest neighbors (inner circle) and second nearest neighbors (outer circle) are identical. 
This demonstrates that the exchange interactions between the first and second neighbors cannot be computed in this supercell using the mapping method, 
as they share the same $\alpha$ coefficient (Eq.~\ref{eq:hes}).
}
\label{fig:hex-2x2}
\end{figure}

To visually demonstrate the limitations of calculating exchange interactions in a supercell, 
we consider a $2 \times 2$ hexagonal monolayer supercell containing 4 magnetic atoms, as shown in Fig.~\ref{fig:hex-2x2}. 
Each atom is assigned a distinct color to represent an independent magnetic direction. 
However, due to periodic boundary conditions, these colors repeat, 
leading to a situation where the first and second nearest neighbors share the same set of distinct colors. 
This implies that the coefficients $\alpha_{k,1}$ and $\alpha_{k,2}$ are identical, 
making it impossible to independently determine the exchange interactions $J_1$ and $J_2$ in this supercell.

In more complex systems, this limitation may not be as easily detected, so a mathematical approach is necessary. 
By applying Gaussian elimination and analyzing the null space of matrix $A$, 
we derive a general relationship: $\alpha_{k,l} = \beta_0 + \sum_{i<l} \beta_i \alpha_{k,i}$, 
which shows that $J_l$ depends on the preceding exchange interactions $J_{i<l}$ due to periodic boundary conditions. 
To calculate exchange interactions beyond the \( l \)-th nearest neighbor, a larger supercell is required.

To calculate exchange interactions up to a specific number of nearest neighbors, 
identifying the smallest supercell that permits the computation of these interactions can significantly reduce computational costs. 
Researchers typically extend the primitive cell by integer factors $n$, $m$, and $q$ to form a conventional $n \times m \times q$ supercell. 
However, generating supercell structures can be more generally achieved using a $3 \times 3$ transformation matrix. 
By applying all possible transformations (up to a specific supercell size), 
one can explore the supercell space and identify the optimal supercell structure for calculating the desired exchange interactions.

In this work, we present a computational method for efficiently identifying supercell structures that meet the desired criteria. 
In the following sections, we provide a detailed explanation of our approach for selecting the optimal supercell structures.  
Next, we demonstrate the computational time savings achieved with this method using specific examples.  
Finally, we present a practical example of applying density functional theory (DFT) with the optimal supercell structures to calculate the exchange parameters of Fe$_2$O$_3$.

{\it Building supercell structures} $ - $
To construct a supercell from a parent cell, we can use an integer matrix ${\mathbb H}$  
to transform the parent lattice vectors $\mathbb{V}_{\mathrm{p}}$ into the supercell lattice vectors
$\mathbb{V}_{\mathrm{s}}$ using $\mathbb{V}_{\mathrm{s}}=\mathbb{H} \mathbb{V}_{\mathrm{p}}$. 
The lattice vectors are represented in rows in both  $\mathbb{V}_{\mathrm{s}}$ and $\mathbb{V}_{\mathrm{p}}$.
To generate all possible supercells for a specific size $n$, 
we use Hermite normal form (HNF) matrices to transform the parent lattice and 
then apply rotational symmetry of the lattice to eliminate duplicate supercells. 
It has been shown that all possible HNF matrices can be created in an upper-triangular form~\cite{Santoro1973,Santoro1972,Gus2008}:

\begin{equation}
{\mathbb{H}}=\begin{bmatrix}
a & b & c \\
0 & d & e \\
0 & 0 & f
\end{bmatrix}
\quad 0 \leq b < d, \quad 0 \leq c, e < f ,
\end{equation}
where $a,..,f$ are integers. In addition to the inequality constraints, 
the size $n$ of the supercell dictates the determinant of the matrix ($\mid {\mathbb {H}} \mid = a \times d \times f=n$).
Using these constraints, all possible combinations of $a, \dots, f$ can be identified~\cite{Gus2008, Gus2009}.

However, relying solely on the HNF matrix transformation 
may result in superlattices with lattice vectors that differ greatly in length, making them impractical. 
Minkowski reduction~\cite{Lattice_reduct1, Lattice_reduct2} helps 
to address this issue by providing shorter, more orthogonal lattice vectors 
without altering the lattice volume. 
Therefore, in this work, 
we apply Minkowski reduction after performing the HNF matrix transformation.

{\it Finding optimal supercell structures} $ - $
For each magnetic configuration (e.g., $k$-th), within a supercell, the total energy of the electronic structure is related to exchange interactions by Eq.~\ref{eq:hes}. 
For $n$ magnetic configurations, we can represent this relationship using the following linear algebra equation ($\mathbb {A J = E}$ ):
\begin{align}
\begin{pmatrix}
1 & \alpha_{1,1} & \alpha_{1,2} & \cdots & \alpha_{1,m} \\
1 & \alpha_{2,1} & \alpha_{2,2} & \cdots & \alpha_{2,m} \\
1 & \alpha_{3,1} & \alpha_{3,2} & \cdots & \alpha_{3,m} \\
\vdots & \vdots & \vdots & \ddots & \vdots \\
1 & \alpha_{n,1} & \alpha_{n,2} & \cdots & \alpha_{n,m}
\end{pmatrix}
&
\begin{pmatrix}
c_0 \\
J_1 \\
J_2 \\
\vdots \\
J_{m}
\end{pmatrix}
	&=&
\begin{pmatrix}
E_1 \\
E_2 \\
E_3 \\
\vdots \\
E_n
\end{pmatrix}
\end{align}
By using Gaussian elimination to find the Reduced Row Echelon Form (RREF) of matrix $\mathbb A$ 
and identifying null space vectors~\cite{linearalg}, we can determine which columns depend on earlier columns. 
This tells us the maximum distance (nearest neighbor) for calculating exchange interactions. 
For example, if the column for the $q$-th nearest neighbor (column $q+1$) is the first one that depends on previous columns, 
we should set $J_{i \geq q} = 0 $ to avoid incorrect results due to this dependency. 
To include exchange interactions beyond the $q$-th nearest neighbor, a larger supercell is needed.
Note that the rank of matrix $A$ may not depend on the the index of the first dependent column.
This occurs because, in some cases, columns following the first dependent column may be linearly independent on the preceding columns. 
As a result, the rank of the matrix can exceed $q$. 
Therefore, focusing solely on rank can be misleading when determining the limitations of a supercell.

To search for the optimal supercell for maximizing the number of exchange interactions within a range of sizes, the following simplified steps can be applied:
\begin{enumerate}
	\item Generate all possible supercell structures using HNF matrices, then apply Minkowski reduction.
	\item Create random magnetic configurations for each supercell and construct the ${\mathbb A}$ matrix.
	\item Use Gaussian elimination to find the null space vectors of matrix ${\mathbb A}$ and identify the first dependent column.
	\item Sort the supercells based on the index of the first dependent column.
\end{enumerate}
We have two options in step two: use random or generate all possible linear magnetic configurations. 
However, the total number of configurations increases exponentially 
as the number of magnetic atoms increases, which makes the random approach more practical. 
When constructing matrix ${\mathbb A}$, we remove duplicate rows, as duplicates can arise in small supercells due to symmetry or repeated configurations. 
It's advisable to avoid using such small supercells, as they reduce the number of unique magnetic configurations required to calculate exchange parameters. 
Therefore, the fraction of distinct configurations can also guide the selection of an optimal supercell.

We have made a code available on GitHub (\url{https://github.com/malaei/SUPERHEX/}) that can generate optimal structures within a specified volume range. 
The code requires only a parent cell (e.g., a primitive cell) to find the best supercell structures. It is written in Python, and is both parallelized and efficient.

{\it Results} $-$
To demonstrate the results of our method, we use a CrCl$_3$ monolayer as an example (see Fig. 2). 
We generate all possible supercell sizes from 1 to 16. Applying the algorithm leads us to 
a supercell size of 8, which allows us to calculate exchange parameters up to the $8$th nearest neighbors. 
In contrast, using conventional supercell construction, we would need a $4\times4$ supercell (size 16) to calculate the same exchange parameters. 
This results in an approximately 8-fold reduction in computational time for methods like DFT, which scale as $n^3$, where $n$ is the number of electrons.

Table~\ref{tab:examples} presents several examples that highlight the method’s efficiency. 
As shown in the table, this approach can speed up calculations by nearly an order of magnitude in many cases.
According to Table~\ref{tab:examples}, the method becomes significantly more efficient when the goal is to calculate exchange interactions at greater distances.

\begin{figure}
\includegraphics[scale=0.070]{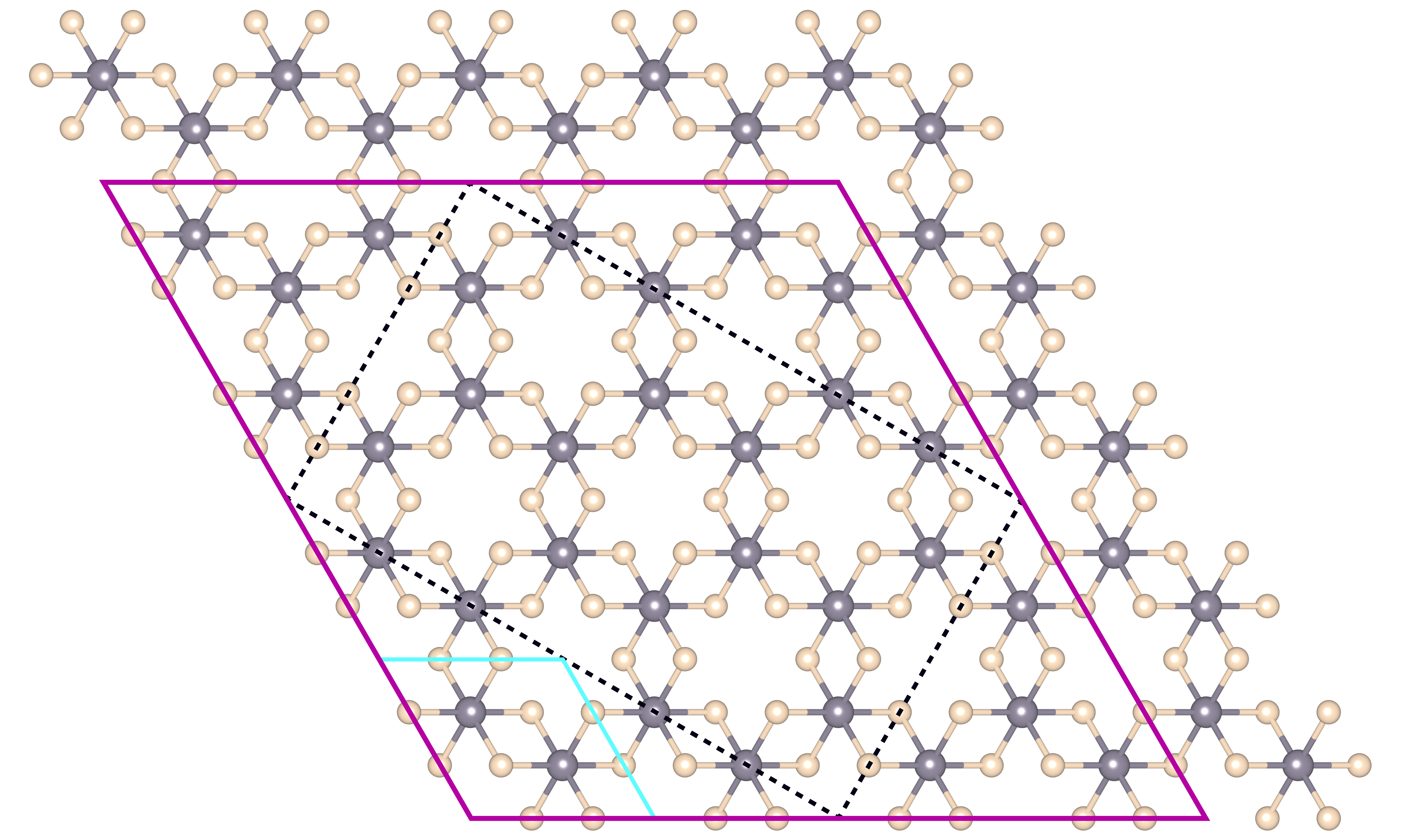}
\caption{
The figures illustrate different cell sizes of CrCl$_3$. 
Larger spheres represent Cr atoms, while smaller spheres represent Cl atoms. 
Cyan lines outline the primitive cell, purple lines indicate a $4 \times 4$ supercell (size 16), and dashed lines mark a supercell of size 8. 
Both supercells allow for calculating the Heisenberg exchange interaction up to the $8$th nearest neighbor. 
However, the larger supercell (size 16) requires approximately 8 times more computational time compared to the supercell of size 8.
	}
\label{fig:CrCl3}
\end{figure}

\begin{figure}
\includegraphics[scale=0.550]{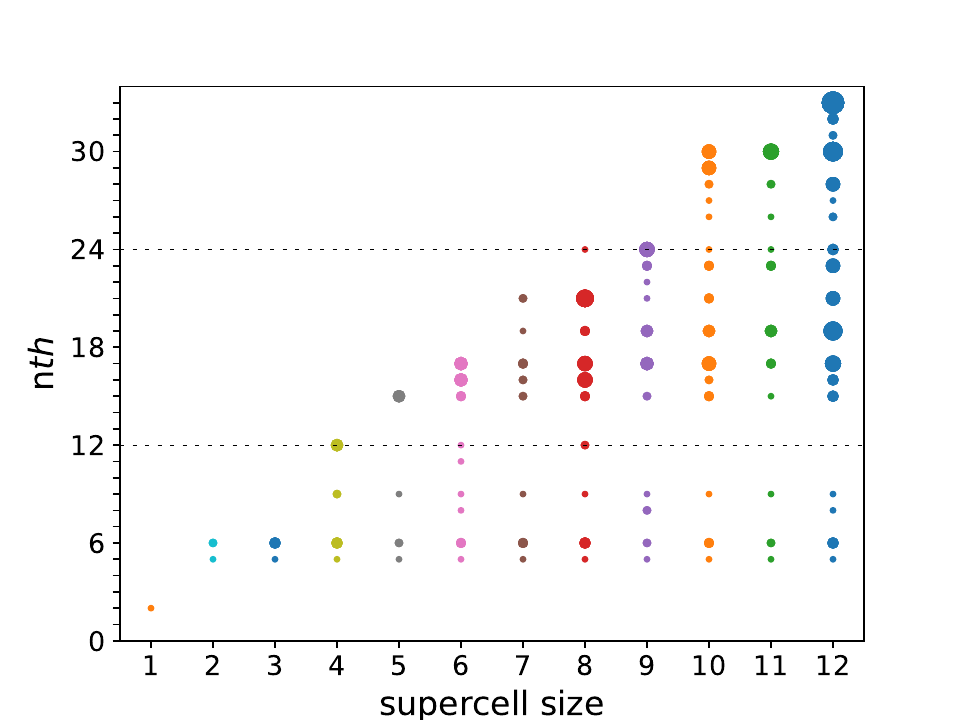}
\caption{
The plot shows the allowed exchange interactions for supercell structures generated with supercell sizes ranging from 1 to 12. 
The x-axis represents different supercell sizes, while the y-axis indicates the number of permitted exchange interactions for each supercell structure. 
For each supercell size, there are multiple distinct supercell structures. For example, at supercell size 8, there are 42 distinct supercell structures. 
The size of the circles in the plot reflects the population of structures that 
allow the calculation of a specific number of exchange interactions for these supercell structures.}
\label{fig:js_vs_vol}
\end{figure}

\begin{table}[h!]
\caption{
The table presents examples of magnetic materials to compare our method for finding the optimal supercell with the conventional method. 
The primitive cells of NiO, MnTe, Fe$_2$O$_3$, and MnO$_2$ contain 2, 2, 4, and 2 magnetic atoms, respectively. 
The table also indicates the number of allowed exchange interactions for each optimal supercell size, along with the equivalent conventional supercell that can be used to calculate the same number of exchange interactions.
}
\label{tab:examples}
\small
\centering
\begin{tabular}{ccccc}
\toprule
\toprule
	Material             & \parbox[c]{2cm}{\centering    conventional \\ supercell size}   &  \parbox[c]{1.5cm}{\centering    permitted \\ $J_n$}    &  \parbox[c]{2cm}{\centering optimal \\ supercell size}   & speedup    \\ \hline
\multirow{2}{*}{NiO}         & 8  ($2\times  2\times 2$)                                       & $J_1 \cdots J_5$     &  5                                                        & $\sim 4$   \\ \cline{2-5} 
                             & 27 ($3\times  3\times 3$)                                       & $J_1 \cdots J_7$     &  7                                                        & $\sim 57$  \\ \hline
\multirow{2}{*}{MnTe}        & 18 ($3\times  3\times 2$)                                       & $J_1 \cdots J_7$     &  8                                                        & $\sim 11$  \\ \cline{2-5} 
		             & 48 ($4\times  4\times 3$)                                       & $J_1 \cdots J_{13}$  &  12                                                       & $\sim 64$  \\ \hline
\multirow{2}{*}{Fe$_2$O$_3$} & 8  ($2\times  2\times 2$)                                       & $J_1 \cdots J_{12}$  &  4                                                        & $\sim 8$   \\ \cline{2-5} 
		             & 27 ($3\times  3\times 3$)                                       & $J_1 \cdots J_{22}$  &  8                                                        & $\sim 38$  \\ 
\bottomrule
\end{tabular}
\end{table}
Amongst the materials listed in Table~\ref{tab:examples}, 
we chose Fe$_2$O$_3$ to showcase the effectiveness of our method in finding an optimal supercell for calculating exchange interactions.
When generating supercell structures with sizes ranging from 1 to 12, we obtain 322 distinct structures for Fe$_2$O$_3$. 
Figure~\ref{fig:js_vs_vol} displays the maximum number of nearest neighbors that can be considered for each supercell structure in an ab initio calculation. 
As shown, different structures allow varying numbers of nearest neighbors for each supercell size. 
Since the supercell size primarily influences the computation time, it is recommended to choose the structure with the highest number of permitted nearest neighbors for more efficient calculations.

\begin{figure}
\includegraphics[scale=0.550]{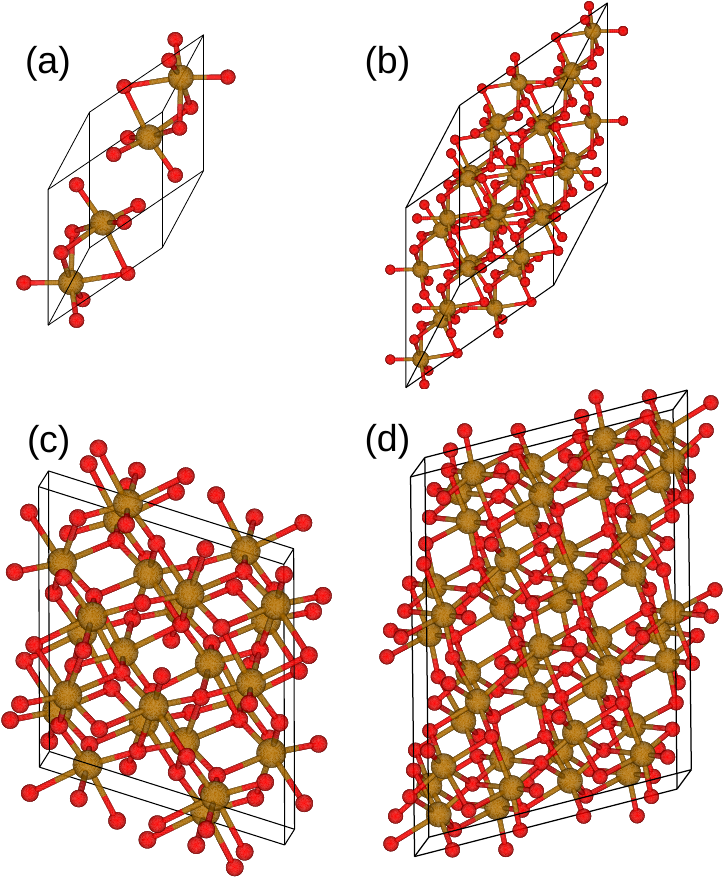}
\caption{
The figure shows the primitive and supercell structures of Fe$_2$O$_3$. 
(a) represents the primitive cell, (b) illustrates the conventional $2 \times 2 \times 2$ supercell, 
(c) displays the supercell structure of size 4, which allows the calculation of exchange interactions up to $J_{12}$, 
and (d) shows the supercell structure of size 8, which permits the calculation of exchange interactions up to $J_{24}$. 
	The small spheres represent oxygen atoms, while the large spheres represent iron atoms.}
\label{fig:Fe2O3}
\end{figure}

\begin{figure}
\includegraphics[scale=0.550]{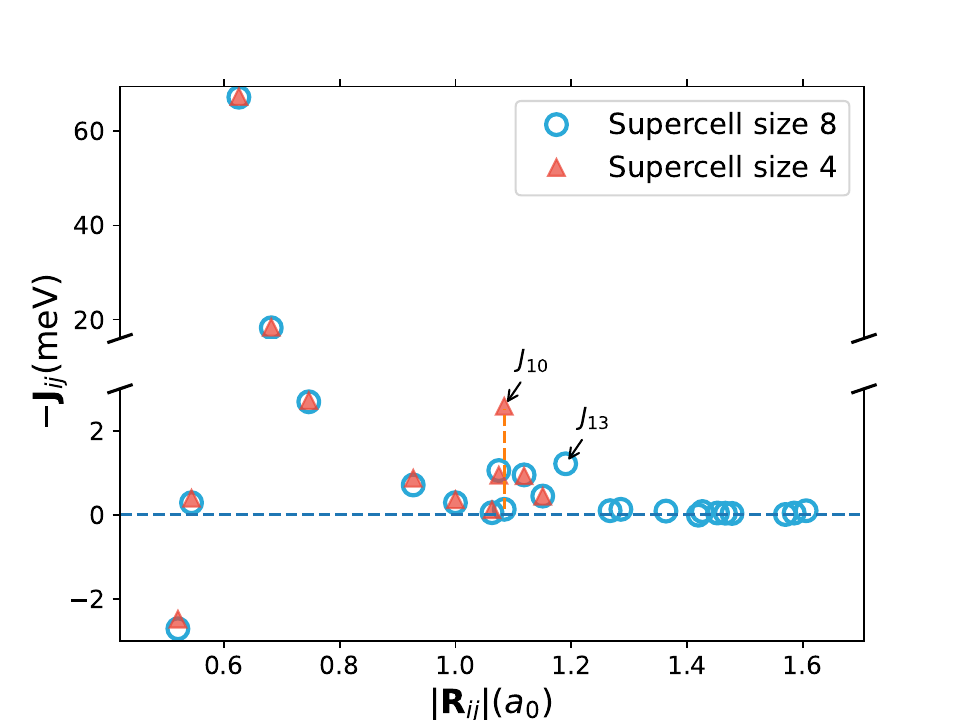}
\caption{
The plot presents DFT results for two supercell structures of Fe$_2$O$_3$, one with a size of 4 and the other with a size of 8. 
Using the structure of size 4, we can calculate Heisenberg exchange interactions up to $J_{12}$. 
However, with a supercell of size 8, we can estimate exchange parameters up to $J_{24}$. }
\label{fig:Fe2O3_Js}
\end{figure}

According to Table~\ref{tab:examples} and the plot in Figure~\ref{fig:js_vs_vol}, a supercell structure of size 4 can be used to compute exchange interactions up to $J_{12}$. 
This structure, along with other Fe$_2$O$_3$ supercell structure, is visualized in Figure~\ref{fig:Fe2O3}. 
Using the least-squares method, we fit DFT results from 48 distinct magnetic configurations within this supercell to the Heisenberg Hamiltonian, 
yielding exchange interactions up to $J_{12}$, as shown in Figure~\ref{fig:Fe2O3_Js}. 
The significant values of $J_{11}$ and $J_{12}$ suggest that interactions beyond $J_{12}$ may be relevant, 
indicating the need for a larger supercell for accurate calculation.

As shown in Figure~\ref{fig:js_vs_vol}, a supercell structure of size 8 (visualized in Figure~\ref{fig:Fe2O3}) 
allows for the calculation of exchange interactions up to $J_{24}$. 
In contrast, a conventional supercell of the same size ($2 \times 2 \times 2$) only permits calculations up to $J_{12}$ (see Table~\ref{tab:examples}). 
According to Table 1, a conventional supercell larger than $3 \times 3 \times 3$ would be necessary to compute interactions up to $J_{24}$, 
underscoring the importance of selecting an optimal supercell structure.

With the new supercell structure of size 8, we observe that the value of $J_{10}$ differs significantly from those obtained using the size 4 supercell. 
Additionally, we find notable values for $J_{13}$ (Figure~\ref{fig:Fe2O3_Js}). 
Beyond the 13th nearest neighbor, the exchange interactions approach zero, suggesting that interactions up to $J_{13}$ are sufficient for accurate modeling. 
Thus, using the size 8 supercell derived from our algorithm, we can compute exchange interactions up to $J_{24}$. 
In contrast, the conventional $2 \times 2 \times 2$ supercell only resolves interactions up to $J_{12}$, 
where ambiguities remain, as $J_{10}$, $J_{11}$, and $J_{12}$ still exhibit significant values, 
suggesting that interactions beyond $J_{12}$ may still play a role.

{\it Conclusion} $ - $
In this work, we introduced an innovative method to identify optimal supercell structures 
for calculating exchange interactions by mapping spin-polarized {\it ab initio} results onto the Heisenberg Hamiltonian. 
This method can significantly accelerate exchange interaction calculations, often by an order of magnitude. 
It is particularly beneficial for {\it ab initio} methods with convergence challenges, such as meta-GGA methods~\cite{SCAN, r2SCAN}. 
By selecting the smallest possible supercell, convergence issues are mitigated, especially in spin-polarized calculations, where these challenges are typically more pronounced.

{\it Acknowledgements} $ - $
M. A. thanks Nafiseh Rezaei for her valuable comments and assistance in improving Figures ~\ref{fig:hex-2x2} and ~\ref{fig:CrCl3}.
\bibliography{refs.bib}

\begin{thebibliography}{17}%
\makeatletter
\providecommand \@ifxundefined [1]{%
 \@ifx{#1\undefined}
}%
\providecommand \@ifnum [1]{%
 \ifnum #1\expandafter \@firstoftwo
 \else \expandafter \@secondoftwo
 \fi
}%
\providecommand \@ifx [1]{%
 \ifx #1\expandafter \@firstoftwo
 \else \expandafter \@secondoftwo
 \fi
}%
\providecommand \natexlab [1]{#1}%
\providecommand \enquote  [1]{``#1''}%
\providecommand \bibnamefont  [1]{#1}%
\providecommand \bibfnamefont [1]{#1}%
\providecommand \citenamefont [1]{#1}%
\providecommand \href@noop [0]{\@secondoftwo}%
\providecommand \href [0]{\begingroup \@sanitize@url \@href}%
\providecommand \@href[1]{\@@startlink{#1}\@@href}%
\providecommand \@@href[1]{\endgroup#1\@@endlink}%
\providecommand \@sanitize@url [0]{\catcode `\\12\catcode `\$12\catcode
  `\&12\catcode `\#12\catcode `\^12\catcode `\_12\catcode `\%12\relax}%
\providecommand \@@startlink[1]{}%
\providecommand \@@endlink[0]{}%
\providecommand \url  [0]{\begingroup\@sanitize@url \@url }%
\providecommand \@url [1]{\endgroup\@href {#1}{\urlprefix }}%
\providecommand \urlprefix  [0]{URL }%
\providecommand \Eprint [0]{\href }%
\providecommand \doibase [0]{https://doi.org/}%
\providecommand \selectlanguage [0]{\@gobble}%
\providecommand \bibinfo  [0]{\@secondoftwo}%
\providecommand \bibfield  [0]{\@secondoftwo}%
\providecommand \translation [1]{[#1]}%
\providecommand \BibitemOpen [0]{}%
\providecommand \bibitemStop [0]{}%
\providecommand \bibitemNoStop [0]{.\EOS\space}%
\providecommand \EOS [0]{\spacefactor3000\relax}%
\providecommand \BibitemShut  [1]{\csname bibitem#1\endcsname}%
\let\auto@bib@innerbib\@empty
\bibitem [{\citenamefont {Mosleh}\ and\ \citenamefont
  {Alaei}(2023)}]{Alaei2023}%
  \BibitemOpen
  \bibfield  {author} {\bibinfo {author} {\bibfnamefont {Z.}~\bibnamefont
  {Mosleh}}\ and\ \bibinfo {author} {\bibfnamefont {M.}~\bibnamefont {Alaei}},\
  }\bibfield  {title} {\bibinfo {title} {Benchmarking density functional theory
  on the prediction of antiferromagnetic transition temperatures},\ }\href
  {https://doi.org/10.1103/PhysRevB.108.144413} {\bibfield  {journal} {\bibinfo
   {journal} {Phys. Rev. B}\ }\textbf {\bibinfo {volume} {108}},\ \bibinfo
  {pages} {144413} (\bibinfo {year} {2023})}\BibitemShut {NoStop}%
\bibitem [{\citenamefont {Needs}\ \emph {et~al.}(2009)\citenamefont {Needs},
  \citenamefont {Towler}, \citenamefont {Drummond},\ and\ \citenamefont
  {Ríos}}]{Needs2010}%
  \BibitemOpen
  \bibfield  {author} {\bibinfo {author} {\bibfnamefont {R.~J.}\ \bibnamefont
  {Needs}}, \bibinfo {author} {\bibfnamefont {M.~D.}\ \bibnamefont {Towler}},
  \bibinfo {author} {\bibfnamefont {N.~D.}\ \bibnamefont {Drummond}},\ and\
  \bibinfo {author} {\bibfnamefont {P.~L.}\ \bibnamefont {Ríos}},\ }\bibfield
  {title} {\bibinfo {title} {Continuum variational and diffusion quantum monte
  carlo calculations},\ }\href {https://doi.org/10.1088/0953-8984/22/2/023201}
  {\bibfield  {journal} {\bibinfo  {journal} {Journal of Physics: Condensed
  Matter}\ }\textbf {\bibinfo {volume} {22}},\ \bibinfo {pages} {023201}
  (\bibinfo {year} {2009})}\BibitemShut {NoStop}%
\bibitem [{\citenamefont {Booth}\ \emph {et~al.}(2013)\citenamefont {Booth},
  \citenamefont {Gr{\"u}neis}, \citenamefont {Kresse},\ and\ \citenamefont
  {Alavi}}]{CCSD_nature}%
  \BibitemOpen
  \bibfield  {author} {\bibinfo {author} {\bibfnamefont {G.~H.}\ \bibnamefont
  {Booth}}, \bibinfo {author} {\bibfnamefont {A.}~\bibnamefont {Gr{\"u}neis}},
  \bibinfo {author} {\bibfnamefont {G.}~\bibnamefont {Kresse}},\ and\ \bibinfo
  {author} {\bibfnamefont {A.}~\bibnamefont {Alavi}},\ }\bibfield  {title}
  {\bibinfo {title} {Towards an exact description of electronic wavefunctions
  in real solids},\ }\href@noop {} {\bibfield  {journal} {\bibinfo  {journal}
  {Nature}\ }\textbf {\bibinfo {volume} {493}},\ \bibinfo {pages} {365}
  (\bibinfo {year} {2013})}\BibitemShut {NoStop}%
\bibitem [{\citenamefont {Čížek}(1966)}]{CCSD}%
  \BibitemOpen
  \bibfield  {author} {\bibinfo {author} {\bibfnamefont {J.}~\bibnamefont
  {Čížek}},\ }\bibfield  {title} {\bibinfo {title} {{On the Correlation
  Problem in Atomic and Molecular Systems. Calculation of Wavefunction
  Components in Ursell‐Type Expansion Using Quantum‐Field Theoretical
  Methods}},\ }\href {https://doi.org/10.1063/1.1727484} {\bibfield  {journal}
  {\bibinfo  {journal} {The Journal of Chemical Physics}\ }\textbf {\bibinfo
  {volume} {45}},\ \bibinfo {pages} {4256} (\bibinfo {year} {1966})},\ \Eprint
  {https://arxiv.org/abs/https://pubs.aip.org/aip/jcp/article-pdf/45/11/4256/18845884/4256\_1\_online.pdf}
  {https://pubs.aip.org/aip/jcp/article-pdf/45/11/4256/18845884/4256\_1\_online.pdf}
  \BibitemShut {NoStop}%
\bibitem [{\citenamefont {Knizia}\ and\ \citenamefont {Chan}(2012)}]{DMET1}%
  \BibitemOpen
  \bibfield  {author} {\bibinfo {author} {\bibfnamefont {G.}~\bibnamefont
  {Knizia}}\ and\ \bibinfo {author} {\bibfnamefont {G.~K.-L.}\ \bibnamefont
  {Chan}},\ }\bibfield  {title} {\bibinfo {title} {Density matrix embedding: A
  simple alternative to dynamical mean-field theory},\ }\href
  {https://doi.org/10.1103/PhysRevLett.109.186404} {\bibfield  {journal}
  {\bibinfo  {journal} {Phys. Rev. Lett.}\ }\textbf {\bibinfo {volume} {109}},\
  \bibinfo {pages} {186404} (\bibinfo {year} {2012})}\BibitemShut {NoStop}%
\bibitem [{\citenamefont {Cui}\ \emph {et~al.}(2019)\citenamefont {Cui},
  \citenamefont {Zhu},\ and\ \citenamefont {Chan}}]{DMET2}%
  \BibitemOpen
  \bibfield  {author} {\bibinfo {author} {\bibfnamefont {Z.-H.}\ \bibnamefont
  {Cui}}, \bibinfo {author} {\bibfnamefont {T.}~\bibnamefont {Zhu}},\ and\
  \bibinfo {author} {\bibfnamefont {G.~K.-L.}\ \bibnamefont {Chan}},\
  }\bibfield  {title} {\bibinfo {title} {Efficient implementation of ab initio
  quantum embedding in periodic systems: Density matrix embedding theory},\
  }\href@noop {} {\bibfield  {journal} {\bibinfo  {journal} {Journal of
  Chemical Theory and Computation}\ }\textbf {\bibinfo {volume} {16}},\
  \bibinfo {pages} {119} (\bibinfo {year} {2019})}\BibitemShut {NoStop}%
\bibitem [{\citenamefont {Szilva}\ \emph {et~al.}(2023)\citenamefont {Szilva},
  \citenamefont {Kvashnin}, \citenamefont {Stepanov}, \citenamefont
  {Nordstr\"om}, \citenamefont {Eriksson}, \citenamefont {Lichtenstein},\ and\
  \citenamefont {Katsnelson}}]{Lichtenstein2023}%
  \BibitemOpen
  \bibfield  {author} {\bibinfo {author} {\bibfnamefont {A.}~\bibnamefont
  {Szilva}}, \bibinfo {author} {\bibfnamefont {Y.}~\bibnamefont {Kvashnin}},
  \bibinfo {author} {\bibfnamefont {E.~A.}\ \bibnamefont {Stepanov}}, \bibinfo
  {author} {\bibfnamefont {L.}~\bibnamefont {Nordstr\"om}}, \bibinfo {author}
  {\bibfnamefont {O.}~\bibnamefont {Eriksson}}, \bibinfo {author}
  {\bibfnamefont {A.~I.}\ \bibnamefont {Lichtenstein}},\ and\ \bibinfo {author}
  {\bibfnamefont {M.~I.}\ \bibnamefont {Katsnelson}},\ }\bibfield  {title}
  {\bibinfo {title} {Quantitative theory of magnetic interactions in solids},\
  }\href {https://doi.org/10.1103/RevModPhys.95.035004} {\bibfield  {journal}
  {\bibinfo  {journal} {Rev. Mod. Phys.}\ }\textbf {\bibinfo {volume} {95}},\
  \bibinfo {pages} {035004} (\bibinfo {year} {2023})}\BibitemShut {NoStop}%
\bibitem [{\citenamefont {Sadeghi}\ \emph {et~al.}(2015)\citenamefont
  {Sadeghi}, \citenamefont {Alaei}, \citenamefont {Shahbazi},\ and\
  \citenamefont {Gingras}}]{Alaei2015}%
  \BibitemOpen
  \bibfield  {author} {\bibinfo {author} {\bibfnamefont {A.}~\bibnamefont
  {Sadeghi}}, \bibinfo {author} {\bibfnamefont {M.}~\bibnamefont {Alaei}},
  \bibinfo {author} {\bibfnamefont {F.}~\bibnamefont {Shahbazi}},\ and\
  \bibinfo {author} {\bibfnamefont {M.~J.~P.}\ \bibnamefont {Gingras}},\
  }\bibfield  {title} {\bibinfo {title} {Spin hamiltonian, order out of a
  coulomb phase, and pseudocriticality in the frustrated pyrochlore heisenberg
  antiferromagnet ${\mathrm{fef}}_{3}$},\ }\href
  {https://doi.org/10.1103/PhysRevB.91.140407} {\bibfield  {journal} {\bibinfo
  {journal} {Phys. Rev. B}\ }\textbf {\bibinfo {volume} {91}},\ \bibinfo
  {pages} {140407} (\bibinfo {year} {2015})}\BibitemShut {NoStop}%
\bibitem [{\citenamefont {Santoro}\ and\ \citenamefont
  {Mighell}(1973)}]{Santoro1973}%
  \BibitemOpen
  \bibfield  {author} {\bibinfo {author} {\bibfnamefont {A.}~\bibnamefont
  {Santoro}}\ and\ \bibinfo {author} {\bibfnamefont {A.~D.}\ \bibnamefont
  {Mighell}},\ }\bibfield  {title} {\bibinfo {title} {{Coincidence-site
  lattices}},\ }\href {https://doi.org/10.1107/S0567739473000434} {\bibfield
  {journal} {\bibinfo  {journal} {Acta Crystallographica Section A}\ }\textbf
  {\bibinfo {volume} {29}},\ \bibinfo {pages} {169} (\bibinfo {year}
  {1973})}\BibitemShut {NoStop}%
\bibitem [{\citenamefont {Santoro}\ and\ \citenamefont
  {Mighell}(1972)}]{Santoro1972}%
  \BibitemOpen
  \bibfield  {author} {\bibinfo {author} {\bibfnamefont {A.}~\bibnamefont
  {Santoro}}\ and\ \bibinfo {author} {\bibfnamefont {A.~D.}\ \bibnamefont
  {Mighell}},\ }\bibfield  {title} {\bibinfo {title} {{Properties of crystal
  lattices: the derivative lattices and their determination}},\ }\href
  {https://doi.org/10.1107/S0567739472000737} {\bibfield  {journal} {\bibinfo
  {journal} {Acta Crystallographica Section A}\ }\textbf {\bibinfo {volume}
  {28}},\ \bibinfo {pages} {284} (\bibinfo {year} {1972})}\BibitemShut
  {NoStop}%
\bibitem [{\citenamefont {Hart}\ and\ \citenamefont {Forcade}(2008)}]{Gus2008}%
  \BibitemOpen
  \bibfield  {author} {\bibinfo {author} {\bibfnamefont {G.~L.~W.}\
  \bibnamefont {Hart}}\ and\ \bibinfo {author} {\bibfnamefont {R.~W.}\
  \bibnamefont {Forcade}},\ }\bibfield  {title} {\bibinfo {title} {Algorithm
  for generating derivative structures},\ }\href
  {https://doi.org/10.1103/PhysRevB.77.224115} {\bibfield  {journal} {\bibinfo
  {journal} {Phys. Rev. B}\ }\textbf {\bibinfo {volume} {77}},\ \bibinfo
  {pages} {224115} (\bibinfo {year} {2008})}\BibitemShut {NoStop}%
\bibitem [{\citenamefont {Hart}\ and\ \citenamefont {Forcade}(2009)}]{Gus2009}%
  \BibitemOpen
  \bibfield  {author} {\bibinfo {author} {\bibfnamefont {G.~L.~W.}\
  \bibnamefont {Hart}}\ and\ \bibinfo {author} {\bibfnamefont {R.~W.}\
  \bibnamefont {Forcade}},\ }\bibfield  {title} {\bibinfo {title} {Generating
  derivative structures from multilattices: Algorithm and application to hcp
  alloys},\ }\href {https://doi.org/10.1103/PhysRevB.80.014120} {\bibfield
  {journal} {\bibinfo  {journal} {Phys. Rev. B}\ }\textbf {\bibinfo {volume}
  {80}},\ \bibinfo {pages} {014120} (\bibinfo {year} {2009})}\BibitemShut
  {NoStop}%
\bibitem [{\citenamefont {Nguyen}\ and\ \citenamefont
  {Stehl\'{e}}(2009)}]{Lattice_reduct1}%
  \BibitemOpen
  \bibfield  {author} {\bibinfo {author} {\bibfnamefont {P.~Q.}\ \bibnamefont
  {Nguyen}}\ and\ \bibinfo {author} {\bibfnamefont {D.}~\bibnamefont
  {Stehl\'{e}}},\ }\bibfield  {title} {\bibinfo {title} {Low-dimensional
  lattice basis reduction revisited},\ }\bibfield  {journal} {\bibinfo
  {journal} {ACM Trans. Algorithms}\ }\textbf {\bibinfo {volume} {5}},\ \href
  {https://doi.org/10.1145/1597036.1597050} {10.1145/1597036.1597050} (\bibinfo
  {year} {2009})\BibitemShut {NoStop}%
\bibitem [{\citenamefont {Bremner}(2011)}]{Lattice_reduct2}%
  \BibitemOpen
  \bibfield  {author} {\bibinfo {author} {\bibfnamefont {M.}~\bibnamefont
  {Bremner}},\ }\href@noop {} {\emph {\bibinfo {title} {Lattice basis
  reduction}}}\ (\bibinfo  {publisher} {CRC Press New York},\ \bibinfo {year}
  {2011})\BibitemShut {NoStop}%
\bibitem [{\citenamefont {Singh}(2013)}]{linearalg}%
  \BibitemOpen
  \bibfield  {author} {\bibinfo {author} {\bibfnamefont {K.}~\bibnamefont
  {Singh}},\ }\href@noop {} {\emph {\bibinfo {title} {Linear algebra: step by
  step}}}\ (\bibinfo  {publisher} {OUP Oxford},\ \bibinfo {year}
  {2013})\BibitemShut {NoStop}%
\bibitem [{\citenamefont {Sun}\ \emph {et~al.}(2015)\citenamefont {Sun},
  \citenamefont {Ruzsinszky},\ and\ \citenamefont {Perdew}}]{SCAN}%
  \BibitemOpen
  \bibfield  {author} {\bibinfo {author} {\bibfnamefont {J.}~\bibnamefont
  {Sun}}, \bibinfo {author} {\bibfnamefont {A.}~\bibnamefont {Ruzsinszky}},\
  and\ \bibinfo {author} {\bibfnamefont {J.~P.}\ \bibnamefont {Perdew}},\
  }\bibfield  {title} {\bibinfo {title} {Strongly constrained and appropriately
  normed semilocal density functional},\ }\href
  {https://doi.org/10.1103/PhysRevLett.115.036402} {\bibfield  {journal}
  {\bibinfo  {journal} {Phys. Rev. Lett.}\ }\textbf {\bibinfo {volume} {115}},\
  \bibinfo {pages} {036402} (\bibinfo {year} {2015})}\BibitemShut {NoStop}%
\bibitem [{\citenamefont {Furness}\ \emph {et~al.}(2020)\citenamefont
  {Furness}, \citenamefont {Kaplan}, \citenamefont {Ning}, \citenamefont
  {Perdew},\ and\ \citenamefont {Sun}}]{r2SCAN}%
  \BibitemOpen
  \bibfield  {author} {\bibinfo {author} {\bibfnamefont {J.~W.}\ \bibnamefont
  {Furness}}, \bibinfo {author} {\bibfnamefont {A.~D.}\ \bibnamefont {Kaplan}},
  \bibinfo {author} {\bibfnamefont {J.}~\bibnamefont {Ning}}, \bibinfo {author}
  {\bibfnamefont {J.~P.}\ \bibnamefont {Perdew}},\ and\ \bibinfo {author}
  {\bibfnamefont {J.}~\bibnamefont {Sun}},\ }\bibfield  {title} {\bibinfo
  {title} {Accurate and numerically efficient r2scan meta-generalized gradient
  approximation},\ }\href@noop {} {\bibfield  {journal} {\bibinfo  {journal}
  {The journal of physical chemistry letters}\ }\textbf {\bibinfo {volume}
  {11}},\ \bibinfo {pages} {8208} (\bibinfo {year} {2020})}\BibitemShut
  {NoStop}%
\end{thebibliography}%

\end{document}